# Moral Geometry: Endogenous Scaling in Nash–Kantian Games

Igor Sloev[1,2], Gerasimos Lianos[3]

## Abstract

We study the strategic implications of the non-invariance of multiplicative Kantian equilibrium (MKE) under monotone transformations of the strategy space. Before interacting with a standard Nash player, a Kantian player publicly selects a smooth increasing scale that determines how proportional deviations are evaluated. Material payoffs and feasible actions remain unchanged, but the chosen scale alters the Kantian first-order condition through endogenous elasticity weights. The representation of actions therefore becomes a commitment device. We characterize the stationary outcomes implementable by a common monotone scale. A sharp dichotomy emerges. Under strategic substitutes, the Kantian player can approach the Nash payoff arbitrarily closely but cannot exceed player 2's Nash benchmark; scaling is defensive and eliminates the payoff loss associated with naive Kantian behavior. Under strategic complements, scaling becomes offensive: the Kantian player can stationary-implement the Stackelberg leader outcome and obtain a payoff strictly above the Nash benchmark. In the canonical Cournot and differentiated Bertrand examples, we explicitly construct scales satisfying the required local elasticity ratios and verify the second-order conditions, so the stationary outcomes are local transformed Nash–Kantian equilibria. Allowing player-specific scales would align the Kantian first-order condition with the Stackelberg condition along the entire reaction curve under complements, but would violate monotonicity under substitutes. This reveals a trade-off between universality and strategic flexibility. The results identify endogenous scaling as a commitment mechanism and connect Kantian optimization to strategic leadership and strategic non-equivalence.

---

[1]School of Computational Social Sciences, European University at St. Petersburg, Russia, ORCID: 0000-0003-0286-1034, sia@eu.spb.ru.

[2]Corresponding author: 6/1A Gagarinskaya st., St. Petersburg, 191187 Russia, sia@eu.spb.ru.

[3]Higher Colleges of Technology, Abu Dhabi College, Baniyas, Abu Dhabi, UAE, glianos@hct.ac.ae

## 1. Introduction

In social dilemmas, Nash equilibrium is generically inefficient. The Kantian approach (Roemer, 2010, 2015, 2019) offers an alternative in which agents evaluate their actions by asking what would happen if everyone acted as they do. In the multiplicative Kantian equilibrium (MKE), a profile is stable if no player would want to multiply all actions by the same positive factor. Under mild conditions, MKE is Pareto efficient. At the same time, MKE has a fundamental feature that distinguishes it from standard Nash analysis: it is not strategically invariant under monotone transformations of the action space (Sher, 2020; Roemer, 2020; Sloev & Lianos, 2026a,b). The set of Kantian equilibria, and the welfare implications attached to them, depend on how strategies are parameterized. This raises a natural question: if the units in which actions are measured are not fixed by the environment, then the choice of parametrization itself becomes economically meaningful. This paper takes that observation seriously and asks what happens when a Kantian player can strategically choose the measurement scale before interacting with a standard Nash player.

The existing literature on Kantian equilibrium has focused mainly on the properties of MKE as a solution concept, its efficiency in symmetric populations, and its evolutionary foundations. The problem of strategic non-equivalence has been identified (Sher, 2020; Roemer 2020), and its implications for symmetric settings have been studied in detail (Sloev & Lianos, 2026a,b). What has not been explored is whether a single Kantian agent can unilaterally use the choice of parametrization as a strategic instrument in an asymmetric Nash–Kantian interaction rather than as a coordination device in a symmetric population. This question is economically relevant because real-world moral behavior often involves choices of units and metrics: how to measure volunteer effort, which carbon metric to adopt, or how to index contribution. Such choices are not always fixed ex ante and may themselves be subject to strategic framing.

We study a two-stage game. In the first stage, the Kantian player publicly and irreversibly chooses a strictly increasing smooth function $f$ that defines the scale on which actions are evaluated. In the second stage, players simultaneously choose actions. The Nash player selects his action in the original strategy space and maximizes his payoff in the standard way. The Kantian player, however, evaluates proportional deviations through a transformed representation of actions defined by a smooth monotone mapping $x = f(y)$. The transformation does not alter the underlying payoff functions or the feasible set of realized actions. It changes only the geometry through which the Kantian player formulates collective deviations. Equivalently, one may represent the game in transformed coordinates $y$, with realized actions given by $x = f(y)$. Under this representation, the Nash player's optimization problem is behaviorally unchanged, whereas the Kantian first-order condition depends explicitly on the elasticity structure induced by $f$. Because the transformation is publicly observed, the Nash player understands how the Kantian player evaluates proportional deviations and adjusts his best response accordingly. The choice of scale therefore affects equilibrium outcomes even though it does not modify material payoffs or the underlying strategic environment.

Rewriting the Kantian first-order condition in terms of original actions yields

$$g(x_1)x_1\partial u_2/\partial x_1(x_1,x_2) + g(x_2)x_2\partial u_2/\partial x_2(x_1,x_2) = 0,$$

where

$$g(x) = d\ln f(y)/d\ln y|_{y=f^{-1}(x)} > 0$$

is the local elasticity of the transformation. The weights $g(x_1)$ and $g(x_2)$ are controlled by the Kantian player through the choice of $f$. As a result, the moral principle becomes state dependent: the Kantian can make restraint tighter or looser at different points of the strategy profile, thereby shaping the equilibrium set. This is the central conceptual observation of the paper: under endogenous scaling, the Kantian imperative becomes a flexible strategic instrument.

Our analysis distinguishes three notions of implementability: stationary outcomes (satisfying the first-order Kantian condition), local equilibria (additionally satisfying the second-order condition), and global equilibria. The main results characterize stationary implementability under a common monotone scale. We prove a sharp dichotomy between the two canonical strategic regimes. Under strategic substitutes, the Kantian player cannot achieve a stationary payoff above player 2's Nash benchmark, but can approach it arbitrarily closely, fully eliminating the free-rider disadvantage. The Nash payoff itself is not attained by any interior stationary outcome. Under strategic complements, we show that the Stackelberg leader outcome is stationary implementable, and the Kantian payoff strictly exceeds the Nash payoff. In the canonical linear Cournot and Bertrand examples we verify the second-order conditions explicitly, upgrading the constructed stationary points to local transformed Kantian–Nash equilibria.

If player-specific transformations are allowed, the Kantian first-order condition can be aligned pointwise with the Stackelberg first-order condition along the entire reaction curve. Under strategic complements this alignment is compatible with strictly increasing scales, whereas under strategic substitutes it would require one of the scales to be decreasing, violating the Kantian principle of universality. This reveals a trade-off between moral universality and strategic flexibility, and shows that the restriction to a common monotone scale is a meaningful behavioral constraint.

The ability to support the Stackelberg leader outcome connects our analysis to the literature on strategic commitment and delegation in oligopoly. Vickers (1985) and Fershtman and Judd (1987) showed that firms can gain a strategic advantage by committing to distorted managerial incentives. In those models, commitment operates through a change in the effective objective function of the decision-maker. Our mechanism is fundamentally different: the Kantian player commits not to a different action or payoff function, but to a different geometry of collective deviations. Endogenous scaling thus emerges as a novel commitment device, distinct from both delegation and standard timing-based leadership.

The economic interpretation of this geometry is straightforward. Scaling can be understood as choosing a metric for effort, contribution, or restraint. In Cournot markets, a Kantian firm that can define the unit of output can avoid being driven out of the market by its own moral rule. In price competition, a Kantian firm that controls how prices are normalized can locally attain the Stackelberg leader position. More broadly, the results apply to public goods, ESG standards, voluntary carbon markets, and other settings in which moral behavior is expressed through quantitative commitments whose measurement is itself a strategic choice. The main message is that moral behavior becomes strategically effective only when the geometry of action aggregation is aligned with incentives.

This paper builds on Sloev & Lianos (2026a,b), who showed the strategic non-equivalence of MKE and demonstrated how parametrization shifts can implement efficient outcomes in symmetric settings. The present paper differs in two respects. First, the transformation is chosen unilaterally by a single Kantian agent in an asymmetric interaction. Second, the transformation is a strictly increasing rescaling rather than a shift of the origin. The endogeneity of the transformation, together with the restriction to a common monotone scale, is the source of the main novelty of the paper and generates the substitutes-versus-complements dichotomy that is absent from the symmetric settings.

## 2. Model

Consider a two-player game. Player 1 is a Nash optimizer (N player) and player 2 is a Kantian optimizer (K player). Both players choose actions from a common interval $X = [0, \infty)$. Player $i$'s payoff is given by a twice continuously differentiable function $u_i(x_1, x_2)$ defined on the interior of $X \times X$.

We impose the following assumptions on the primitives of the game.

(A1) Strict concavity in own action. For each fixed $x_j > 0, \partial^2 u_i/\partial x_i^2 < 0$. Hence any critical point of $u_i(\cdot, x_j)$ is a strict local maximum.

(A2) One-sided payoff externalities. The marginal externalities $\partial_2 u_1$ and $\partial_1 u_2$ have the same constant sign on $(0,\infty)^2$. Both strictly positive and strictly negative externalities are allowed.

(A3) Unique interior best response on a relevant interval. There exists an interval $Y \subseteq [0,\infty)$ that contains the Nash equilibrium action $x_2^N$ and the Stackelberg action $x_2^{St}$ (hence the entire segment between them), such that for each $x_j \in Y$ the problem $max_{x_i \geq 0} u_i(x_i, x_j)$ has a unique interior solution $x_i = R_i(x_j) > 0$, characterised by $\partial u_i/\partial x_i = 0$. By the implicit function theorem, the reaction function $R_i$ is continuously differentiable on $Y$.

(A4) Unique interior Nash equilibrium. The system $\partial u_i/\partial x_i = 0, i = 1,2$, has a unique solution $(x_1^N, x_2^N)$ with $x_1^N > 0, x_2^N > 0$.We denote player 2's Nash payoff by $u_2^N = u_2(x_1^N, x_2^N)$.

(A5) Regular crossing of the Kantian player's own first-order condition. Along the Nash reaction curve, the function $B(x_2) = \partial_2 u_2(R_1(x_2), x_2)$ satisfies $B(x_2^N) = 0$ and crosses zero from above:

$B(x_2) > 0$ for $x_2 < x_2^N$, $B(x_2) < 0$ for $x_2 > x_2^N$

in a neighbourhood of $x_2^N$ (a sufficient differentiable condition is $B'(x_2^N) < 0$).

(A6) Unique interior Stackelberg solution. The leader problem $max_{x_2} u_2(R_1(x_2), x_2)$ admits a unique interior maximiser $x_2^{St} > 0$, and the first-order condition reads

$$\partial_1 u_2(R_1(x_2^{St}), x_2^{St}) R_1'(x_2^{St}) + \partial_2 u_2(R_1(x_2^{St}), x_2^{St}) = 0.$$

We denote the follower action by $x_1^{St} = R_1(x_2^{St})$ and player 2's Stackelberg payoff by $u_2^{St} = u_2(x_1^{St}, x_2^{St})$.

For later use we record the slope of player 1's reaction function,

$$R'(x_2) =- (\partial^2 u_1/\partial x_1 \partial x_2)/(\partial^2 u_1/\partial x_1^2).$$

We say the game exhibits strategic substitutes if $R'(x_2) < 0$ and strategic complements if $R'(x_2) > 0$.

### 3. Mixed Nash–Kantian Interaction and Endogenous Scaling

### 3.1. Nash and Kantian players in natural coordinates

Player 1 is the Nash optimizer. Taking $x_2$ as given, he maximises $u_1(x_1, x_2)$. By (A1) and (A3), his unique interior best response is characterised by the first-order condition

$$\partial_1 u_1(x_1, x_2) = 0 \Leftrightarrow x_1 = R(x_2). \tag{3.1}$$

The reaction function $R$ is continuously differentiable on the relevant interval $Y$.

Player 2 is the Kantian optimizer. In the absence of any transformation of the strategy space, he applies the multiplicative Kantian equilibrium in the natural coordinates. According to the MKE concept, a profile $(x_1, x_2)$ is Kantian-stable if $a = 1$ is a global maximiser of $a \mapsto u_2(ax_1, ax_2)$. The first-order condition for this problem is

$$d/da\, u_2(ax_1, ax_2)|_{a=1} = 0,$$

which simplifies to

$$x_1 \partial_1 u_2(x_1, x_2) + x_2 \partial_2 u_2(x_1, x_2) = 0. \tag{3.2}$$

Together, (3.1) and (3.2) define the mixed Nash–Kantian equilibrium in natural coordinates.

To illustrate the analysis, we introduce two canonical parametric examples that will serve as running illustrations throughout the paper.

Cournot duopoly. Let $u_i(x_i, x_j) = \left(a - b(x_i + x_j)\right) x_i - cx_i$, with $a > c \geq 0$ and $b > 0$. Set $M = a - c$. The Nash reaction is $x_1 = (M - bx_2)/(2b)$. The unique interior Nash equilibrium (assured by (A4)) is $x^N = M/(3b)$, and player 2's Nash profit is $u_2^N = M^2/(9b)$.

Bertrand duopoly with differentiated products. Let $u_i(p_i, p_j) = p_i(\alpha - \beta p_i + \gamma p_j)$, with $\alpha, \beta > 0$ and $0 < \gamma < \beta$. For the numerical specification $\alpha = 10, \beta = 1, \gamma = 0.5$, the Nash reaction is $p_1 = 5 + 0.25p_2$. The unique interior Nash equilibrium is $p^N = 20/3 \approx 6.667$, and player 2's Nash profit is $\pi_2^N = 400/9 \approx 44.444$.

### 3.2. The failure of the naive Kantian

The system formed by (3.1) and (3.2) is often unfavorable to the Kantian player.

In the Cournot example, the mixed Nash–Kantian system has no strictly positive interior solution. The only non-negative solution is the corner $x_2 = 0, x_1 = M/(2b)$, which is the monopoly output of the Nash player. At this point the Kantian player earns zero profit. Thus, in quantity competition, a naive Kantian is driven out of the market.

In the Bertrand example, a positive equilibrium exists, but it yields for the Kantian player $p_2 \approx 8.571$ and $\pi_2 \approx 42.857$, which is strictly below the Nash profit of 44.444.

These examples show that applying MKE in the natural coordinates can be disadvantageous for the Kantian player.

### 3.3. The transformation game

We now allow the Kantian player to choose the measurement scale before the game. The interaction proceeds in two stages. The two-stage structure resembles models of strategic commitment and delegation in oligopoly (Vickers, 1985; Fershtman & Judd, 1987). In those models, a first-stage choice (e.g., managerial incentives) alters the own best response of the committing player and thereby shifts the resulting equilibrium. The mechanism here is similar: the scale $f$ affects the Kantian first-order condition—the K-player's own reaction—while the Nash player continues to follow his unchanged best reply $R(x_2)$. Thus, the choice of scale serves as a commitment device, analogous to delegation, but without modifying material payoffs or relying on contractual arrangements.

The mechanism also differs from models of moral preferences, such as Alger and Weibull's (2013) Homo moralis, where agents maximize a convex combination of selfish and Kantian objectives. In our setup, the Kantian player leaves material payoffs unchanged and instead modifies the geometry through which collective deviations are evaluated. Morality operates not through preferences but through the structure of admissible proportional changes.

Stage 1 (choice of scale). The K player publicly and irreversibly chooses a strictly increasing smooth function $f$ with $f(0) = 0$. This function determines the transformed coordinates $y_i$, with realized actions given by $x_i = f(y_i), i = 1,2$.

Stage 2 (choice of actions). Players simultaneously choose $y_1$ and $y_2$. Since $f$ is strictly increasing and publicly known, the Nash player's optimization in y-space is equivalent to optimization in the original x-space, and his reaction function remains $x_1 = R(x_2)$. The Kantian player, however, applies MKE in y-space and therefore evaluates proportional changes in the transformed coordinates:

$$d/da\, u_2\big(f(ay_1), f(ay_2)\big)|_{a=1} = 0.$$

The next subsection translates this condition back into the original coordinates.

### 3.4. Transformed Kantian condition

Define the elasticity of the transformation by

$$g(x) = d\ln f(y)/d\ln y|_{y=f^{-1}(x)} = \left(yf'(y)/f(y)\right)|_{y=f^{-1}(x)} > 0.$$

**Proposition 1 (Transformed Kantian condition).** *Under a strictly increasing smooth transformation $f$ with $f(0) = 0$, the Kantian first-order condition in the transformed game is equivalent to*

$$g(x_1)x_1\partial_1 u_2(x_1, x_2) + g(x_2)x_2\partial_2 u_2(x_1, x_2) = 0. \quad (3.3)$$

*Proof.* Using the chain rule,

$$\begin{aligned} &d/da\, u_2\left(f(ay_1), f(ay_2)\right)|_{a=1} = \\ &= \partial_1 u_2(x_1, x_2)f'(y_1)y_1 + \partial_2 u_2(x_1, x_2)f'(y_2)y_2 \\ &= (x_1\partial_1 u_2)(y_1 f'(y_1)/x_1) + (x_2\partial_2 u_2)(y_2 f'(y_2)/x_2) \\ &= g(x_1)x_1\partial_1 u_2(x_1, x_2) + g(x_2)x_2\partial_2 u_2(x_1, x_2). \end{aligned}$$

Setting this expression to zero gives (3.3). ■

Equation (3.3) is the first-order necessary condition for a stationary point of the Kantian player's objective in y-space.

### 3.5. Equilibrium concepts for the two-stage game

Fix a smooth strictly increasing transformation $f$ with $f(0) = 0$. Write $x_i = f(y_i)$. We distinguish three notions of implementability.

1. Stationary transformed Nash–Kantian outcome. A profile $(x_1, x_2)$ is a stationary outcome under $f$ if (i) the Nash player is at a best response: $x_1 = R(x_2)$ (equation (3.1)), (ii) the Kantian player's first-order condition (3.3) holds.

2. Local transformed Nash–Kantian equilibrium. A stationary outcome is a local equilibrium under $f$ if, in addition, the second-order condition holds:

$$d^2/da^2\, u_2\big(f(ay_1), f(ay_2)\big)|_{a=1} < 0. \tag{3.5}$$

Thus $a = 1$ is a strict local maximum of the Kantian objective along proportional deviations in transformed coordinates.

3. Global transformed Nash–Kantian equilibrium. A stationary outcome is a global equilibrium under $f$ if

$$u_2\big(f(y_1), f(y_2)\big) \geq u_2\big(f(ay_1), f(ay_2)\big)$$

for all admissible $a > 0$.

The rest of the paper focuses on stationary and local implementability. In the canonical Cournot and Bertrand examples we verify the second-order condition explicitly, showing that the constructed stationary points are indeed local transformed Kantian–Nash equilibria.

## 4. Results: Substitutes versus Complements

We now characterize the stationary outcomes that the Kantian player can attain through a common monotone transformation $f$. The analysis relies on a technical lemma which guarantees that any prescribed pair of elasticity values at two distinct points can be realized by a suitable scale, and that the scale can be chosen with vanishing second derivative at those points, thereby simplifying the verification of the second-order condition in the examples.

**Lemma 1 (Elasticity interpolation with zero curvature).**

*Let $x_a, x_b > 0$ be two distinct interior points and let $g_a, g_b > 0$ be arbitrary positive numbers. Then there exists a function $f: [0, \infty) \to [0, \infty)$, continuous on $[0, \infty)$ and $C^2$ on $(0, \infty)$, strictly increasing, with $f(0) = 0$ and $f'(y) > 0$ for all $y > 0$, such that for some preimages $y_a, y_b > 0$, the following holds:*

*(i) $f(y_a) = x_a, f(y_b) = x_b$,*

*(ii) $\big(y_a f'(y_a)\big)/x_a = g_a$, $\big(y_b f'(y_b)\big)/x_b = g_b$,*

*(iii) $f''(y_a) = f''(y_b) = 0$.*

*So any prescribed local elasticity ratio can be achieved.*

*Proof.*
Without loss of generality, suppose $x_a < x_b$ (if not, interchange the labels). The idea is to construct the derivative $p = f'$ as a positive smooth function on an interval $[y_a, y_b]$ with prescribed values $p(y_a) = s_a, p(y_b) = s_b$ and $p'(y_a) = p'(y_b) = 0$, and with total integral equal to $x_b - x_a$.

Choose a large number $Y > 0$ and set the preimages as $y_a = Y, y_b = Y + 1$. The required endpoint slopes are

$$s_a = (g_a x_a)/y_a, s_b = (g_b x_b)/y_b.$$

By taking $Y$ sufficiently large, both $s_a$ and $s_b$ become arbitrarily small. Let $h: [y_a, y_b] \to [0,1]$ be a smooth increasing function with $h(y_a) = 0, h(y_b) = 1$, and $h'(y_a) = h'(y_b) = 0$. Define

$$p_0(y) = \big(1 - h(y)\big)s_a + h(y)s_b.$$

Then $p_0 > 0$ everywhere, $p_0(y_a) = s_a, p_0(y_b) = s_b$, and $p_0'(y_a) = p_0'(y_b) = 0$. Because $s_a, s_b \to 0$ as $Y \to \infty$, we can choose $Y$ so large that

$$\int_{y_a}^{y_b} p_0(t)dt < x_b - x_a.$$

Pick a non-negative smooth function $\beta \in C_c^\infty\big((y_a, y_b)\big)$, not identically zero, and let $M > 0$ be the unique number satisfying

$$\int_{y_a}^{y_b} \big(p_0(t) + M\beta(t)\big)dt = x_b - x_a.$$

Now set $p(y) = p_0(y) + M\beta(y)$. By construction, $p > 0$ on $[y_a, y_b]$, $p(y_a) = s_a, p(y_b) = s_b, p'(y_a) = p'(y_b) = 0$, and $\int_{y_a}^{y_b} p = x_b - x_a$.

Define $f$ on $[y_a, y_b]$ by

$$f(y) = x_a + \int_{y_a}^{y} p(t)dt.$$

Then $f(y_a) = x_a, f(y_b) = x_b, f'(y_i) = s_i$, and $f''(y_i) = p'(y_i) = 0$ for $i = a, b$.

It remains to extend $f$ to the whole half-line while keeping it strictly increasing, $C^2$ on $(0, \infty)$, and continuous at 0 with $f(0) = 0$. In a small left neighbourhood of $y_a$ we let $f$ be affine with slope $s_a$; similarly, in a small right neighbourhood of $y_b$ we let $f$ be affine with slope $s_b$. Because the values, first derivatives, and second derivatives match at the endpoints, the resulting function is $C^2$ on $(0, \infty)$. We then smoothly connect the affine part near $y_a$ to the origin by any strictly increasing $C^2$ curve that starts at (0,0) and ends at the prescribed point and slope; such a curve always exists by a standard Hermite interpolation on a compact interval. The part beyond $y_b$ is extended linearly with slope $s_b$. The final function f satisfies all the required properties.

Finally, by definition of the slopes, $(y_a f'(y_a)/x_a = (y_a s_a)/x_a = g_a$ and similarly $(y_b f'(y_b))/x_b = g_b$, so the elasticity takes the prescribed values. ■

Along the Nash reaction curve we define

$A(x_2) \equiv \partial_1 u_2(R(x_2), x_2)$, $B(x_2) \equiv \partial_2 u_2(R(x_2), x_2)$, $\pi(x_2) \equiv u_2(R(x_2), x_2)$.

The stationary condition (3.3) can then be rewritten on the reaction curve as

$g(R(x_2))R(x_2)A(x_2) + g(x_2)x_2B(x_2) = 0.$

Any point on the reaction curve with $x_1 = R(x_2) > 0$ and $x_2 > 0$ is attainable as a stationary outcome only if $A(x_2)B(x_2) < 0$. If this condition fails — for instance, at the Nash equilibrium where $B(x_2^N) = 0$ — the point cannot be attained by any choice of a strictly increasing scale $f$.

**4.1. Strategic substitutes**

Assume $R'(x_2) < 0$. By (A2), the sign of $A(x_2)$ is constant on the relevant interval.

At the Nash equilibrium, $B(x_2^N) = 0$. By (A5) we have $B(x_2) > 0$ for $x_2 < x_2^N$ and $B(x_2) < 0$ for $x_2 > x_2^N$ in a neighbourhood of $x_2^N$. Hence the attainable set is if $A(x_2) < 0$, then $x_2 \leq x_2^N$;

if $A(x_2) > 0$, then $x_2 \geq x_2^N$.

**Proposition 2 (Stationary defensive implementability under substitutes).**

*Let $R'(x_2) < 0$. For every $\varepsilon > 0$ there exists a strictly increasing $C^2$ transformation $f$ with $f(0) = 0$ such that the corresponding stationary outcome gives the Kantian player a payoff of at least $u_2^N - \varepsilon$.*

*Proof.*
Consider the case $A(x_2) < 0$; the opposite sign is symmetric and will be outlined afterwards. The attainable set is $x_2 \leq x_2^N$. On $(0, x_2^N)$ we have $B > 0$ and $R' < 0$, hence

$\pi'(x_2) = AR' + B > 0.$

Thus $\pi$ is strictly increasing on $(0, x_2^N)$ and $sup\pi = lim_{x_2 \to x_2^N} \pi(x_2) = u_2^N$. Pick $x_2^\varepsilon = x_2^N - \delta$ with $\delta > 0$ small enough so that $\pi(x_2^\varepsilon) \geq u_2^N - \varepsilon$. The required elasticity ratio is

$$g\left(R(x_2^\varepsilon)\right)/g(x_2^\varepsilon) =-\left(x_2^\varepsilon B(x_2^\varepsilon)\right)/\left(R(x_2^\varepsilon)A(x_2^\varepsilon)\right) > 0.$$

By Lemma 1 (with $x_a = R(x_2^\varepsilon), x_b = x_2^\varepsilon$ and $g_a/g_b$ equal to the required ratio), there exists a strictly increasing $C^2$ function $f$ with $f(0) = 0$ whose elasticity attains this ratio. Hence $(R(x_2^\varepsilon),\ x_2^\varepsilon)$ satisfies (3.1) and (3.3) and is a stationary transformed Nash–Kantian outcome. The Kantian player obtains a payoff of at least $u_2^N - \varepsilon$.

If $A(x_2) > 0$, the attainable set is $x_2 \geq x_2^N$. On $(x_2^N, \infty)$ we have $B < 0$ and $R' < 0$, hence $\pi'(x_2) = AR' + B < 0$, so $\pi$ is strictly decreasing and the supremum is again attained at the limit $x_2 \to x_2^N$. Choosing $x_2^\varepsilon = x_2^N + \delta$ gives the same result. ■

**Example 1 (Cournot, negative externality).**

Let $u_2(x_1, x_2) = (10 - x_1 - x_2)x_2$. Then $R(x_2) = (10 - x_2)/2$, $x_2^N = 10/3 \approx 3.333$, $u_2^N = 100/9 \approx 11.111$.

Take $\delta \in (0,10/3)$ and set $x_2^\delta = x_2^N - \delta = 10/3 - \delta$. Then $x_1^\delta = R\left(x_2^\delta\right) = 10/3 + \delta/2$, $A\left(x_2^\delta\right) = -x_2^\delta$, $B\left(x_2^\delta\right) = (3/2)\delta$.

The required elasticity ratio is $\rho_\delta = g\left(x_1^\delta\right)/g\left(x_2^\delta\right) =-\left(x_2^\delta B\left(x_2^\delta\right)\right) / \left(x_1^\delta A\left(x_2^\delta\right)\right)$

$$= B\left(x_2^\delta\right)/x_1^\delta = \left((3/2)\delta\right)/(10/3 + \delta/2) = 9\delta/(20 + 3\delta).$$

By Lemma 1, we can choose a transformation $f_\delta$ with this ratio and with $f_\delta''\left(y_i^\delta\right) = 0$ at the two preimages. The payoff under this scale is

$$\pi\left(x_2^\delta\right) = \left(5 - x_2^\delta/2\right)x_2^\delta = 100/9 - (5/3)\delta - \delta^2/2.$$

Since $(5/3)\delta + \delta^2/2 \to 0$ as $\delta \to 0$, for any $\varepsilon > 0$ we can pick $\delta$ small enough so that $\pi\left(x_2^\delta\right) \geq u_2^N - \varepsilon$.

Now verify the second-order condition. Because $f_\delta''\left(y_i^\delta\right) = 0$, the left-hand side of (3.5) reduces to $2u_{12}q_1q_2 + u_{22}q_2^2$, where $u_{12} = -1, u_{22} = -2$, and $q_i = f'\left(y_i^\delta\right)y_i^\delta = g\left(x_i^\delta\right)x_i^\delta > 0$. Hence $-2q_1q_2 - 2q_2^2 < 0$.

Thus the stationary outcome is a local transformed Nash–Kantian equilibrium. The Nash payoff itself is not attained by any interior stationary outcome (that would require $\rho = 0$, impossible for positive elasticities), but the Kantian player can secure any payoff arbitrarily close to it from below.

### 4.2. Strategic complements

Now assume $R'(x_2) > 0$. Again the sign of A is constant.

The sign pattern of $B$ is the same as before: $B(x_2^N) = 0, B > 0$ for $x_2 < x_2^N$, $B < 0$ for $x_2 > x_2^N$. The attainable set is if $A > 0$, then $x_2 \geq x_2^N$;

if $A < 0$, then $x_2 \leq x_2^N$.

**Proposition 3 (Stationary Stackelberg implementability under complements).**

*Let $R'(x_2) > 0$. There exists a strictly increasing $C^2$ transformation $f$ with $f(0) = 0$ such that the corresponding stationary outcome coincides with the Stackelberg leader solution, and the Kantian player's payoff strictly exceeds $u_2^N$.*

*Proof.*

Consider $A > 0$; the opposite case is symmetric. The attainable set is $x_2 \geq x_2^N$. At the Nash point, $\pi'(x_2^N) = AR' > 0$. By (A6) the leader problem has a unique interior maximum $x_2^{St} > x_2^N$. Hence $\pi'(x_2^{St}) = 0$, i.e.

$$A(x_2^{St})R'(x_2^{St}) + B(x_2^{St}) = 0. \qquad \text{(St)}$$

The required elasticity ratio at the Stackelberg point is obtained from (3.3) by substituting the stationary condition:

$$g\left(R(x_2^{St})\right)/g(x_2^{St}) = -\left(x_2^{St}B(x_2^{St})\right)/\left(R(x_2^{St})A(x_2^{St})\right).$$

Using (St) to replace $B(x_2^{St}) = -A(x_2^{St})R'(x_2^{St})$, we obtain

$$g\left(R(x_2^{St})\right)/g(x_2^{St}) = \left(x_2^{St}/R(x_2^{St})\right)R'(x_2^{St}) > 0.$$

By Lemma 1, there exists a strictly increasing $C^2$ function $f$ with $f(0) = 0$ whose elasticity attains this ratio. Hence $(R(x_2^{St}), x_2^{St})$ satisfies (3.1) and (3.3) and is a stationary outcome. Because $\pi'(x_2^N) > 0$ and $x_2^{St}$ is the unique interior maximizer of $\pi$, we have $\pi(x_2^{St}) > \pi(x_2^N) = u_2^N$. ■

*Remark.* The construction above requires the Stackelberg point to be regular, that is, $R(x_2^{St}) \neq x_2^{St}$. If $R(x_2^{St}) = x_2^{St}$, a common scale can assign only one elasticity value to that action, so the outcome can be implemented only when the required elasticity ratio equals 1. In the canonical examples (Cournot, Bertrand) the Stackelberg point is always regular.

**Example 2 (Bertrand, positive externality).**

Let $u_2(p_1, p_2) = p_2(10 - p_2 + 0.5p_1)$. Then $R(p_2) = 5 + 0.25p_2$, $p_2^N = 20/3 \approx 6.667$, $u_2^N = 400/9 \approx 44.444$.

The Stackelberg leader solution is $p_2^{St} = 50/7 \approx 7.143$, $p_1^{St} = R(p_2^{St}) = 95/14 \approx 6.786$, $u_2^{St} = 3125/70 \approx 44.643 > u_2^N$.

At this point, $A = 0.5p_2^{St} = 25/7$, $B = 10 - 2p_2^{St} + 0.5p_1^{St} = -25/28$, and the required elasticity ratio is $g(p_1^{St})/g(p_2^{St}) =- (p_2^{St}B)/(p_1^{St}A) = 5/19 \approx 0.263$.

By Lemma 1 we can choose a locally affine scale $f$ with this ratio, so that $f''\left(y_i^{St}\right) = 0$. The left-hand side of the second-order condition (3.5) then becomes $2u_{12}q_1q_2 + u_{22}q_2^2$, where $u_{12} = 0.5, u_{22} = -2$, and $q_i = g\left(p_i^{St}\right)p_i^{St}$. Hence condition (3.5) reduces to $q_1q_2 - 2q_2^2 = q_2(q_1 - 2q_2)$. Using the computed elasticity ratio, $q_1/q_2 = (g(p_1^{St})p_1^{St})/(g(p_2^{St})p_2^{St}) = 1/4$. Thus $q_1 - 2q_2 < 0$, so condition (3.5) holds strictly, and the stationary outcome is a local transformed Nash–Kantian equilibrium. The Kantian firm achieves the Stackelberg leader payoff, strictly above the Nash benchmark, by selecting an appropriate scale whose elasticity ratio at the target point equals 5/19.

### 4.3. Beyond common scaling: the universality trade-off

Suppose now that player 2 may choose separate transformations for the two players, $x_i = f_i(y_i)$, with elasticities $g_i(x_i) > 0$. The Kantian first-order condition becomes

$$g_1(x_1)x_1\partial_1u_2(x_1, x_2) + g_2(x_2)x_2\partial_2u_2(x_1, x_2) = 0. \tag{4.1}$$

If one wants (4.1) to coincide pointwise with the Stackelberg first-order condition along the entire reaction curve $x_1 = R(x_2)$, the required ratio is

$$g_1(x_1)/g_2(x_2) = (x_2/x_1)R'(x_2), x_1 = R(x_2). \tag{4.2}$$

**Proposition 4 (First-order alignment with player-specific transformations).**

*With player-specific transformations, the Kantian first-order condition can be made pointwise identical to the Stackelberg first-order condition along the reaction curve if and only if $R'(x_2) > 0$ on that segment. Under strategic complements the right-hand side of (4.2) is positive, hence the required elasticities are positive and compatible with strictly*

*increasing transformations. Under strategic substitutes $R'(x_2) < 0$, so exact alignment would require one of the elasticities to be negative, which would violate monotonicity.*

The construction is straightforward: set $g_2(x_2) \equiv 1$ (which corresponds to the identity transformation $f_2(y) = y$) and define $g_1(x_1)$ from (4.2) using the inverse reaction function $x_2 = R^{-1}(x_1)$ (since $R'$ has constant nonzero sign on the relevant segment, $R$ is strictly monotonic and hence invertible). The resulting $g_1$ is positive whenever $R' > 0$, and the differential equation $yf_1'(y)/f_1(y) = g_1(f_1(y))$ yields a strictly increasing $f_1$. Thus, under strategic complements, the required player-specific transformations exist and can be written explicitly.

This result highlights that the restriction to a common, monotone scale imposes genuine constraints on implementability. Relaxing universality dramatically expands the set of first-order alignable outcomes, but at the cost of abandoning the Kantian principle that the same moral rule applies to all agents.

## 5. Discussion

In standard Nash analysis, monotone transformations of the strategy space are strategically irrelevant because they preserve unilateral best responses (Mas-Colell, Whinston and Green, 1995). Our analysis shows that this invariance breaks down under multiplicative Kantian optimization, where equilibrium conditions are defined through collective proportional deviations. In Kantian games, the parametrization of actions enters directly into the first-order condition through the elasticity weights $g(x_i)$. As a result, the representation of strategies is no longer behaviorally neutral. The "units" in which actions are expressed become payoff-relevant strategic objects.

This observation has broader conceptual implications. Standard game theory typically treats the geometry of the strategy space as exogenous and representationally innocuous. By contrast, our results suggest that equilibrium concepts based on coordinated deviations may require a richer notion of strategic representation. In Kantian optimization, outcomes depend not only on preferences and feasible actions, but also on how proportional deviations are defined and perceived. The failure of strategic equivalence is therefore not a technical anomaly, but a structural feature of collective optimization.

The paper further shows that endogenous representation can function as a commitment technology. The idea that pre-commitment can generate strategic advantage is central to game theory (Schelling, 1980). Under strategic complements, the Kantian player can implement the Stackelberg leader outcome through an appropriate choice of scale, without delegation, timing advantages, or explicit contractual commitment. The mechanism operates through endogenous moral geometry: by modifying the relative weight attached to proportional deviations, the Kantian player commits to a more aggressive equilibrium action, which the Nash player optimally accommodates. This establishes a novel connection between Kantian optimization and the strategic commitment literature. In the classical delegation framework (Vickers, 1985; Fershtman and Judd, 1987), firms strategically alter managerial incentives to influence rivals' responses. Here, strategic influence emerges not from changing objectives, but from changing the geometry through which collective deviations are evaluated.

Existing approaches to moral behavior in games typically modify preferences. Alger and Weibull (2013) introduce Homo moralis, an agent whose utility is a convex combination of selfish and Kantian objectives. More generally, the literature on social preferences studies how altruism, reciprocity, and fairness concerns reshape equilibrium outcomes. By contrast, our paper leaves material payoffs unchanged and instead alters the geometry of collective deviations. Morality operates not through preferences but through the structure of admissible proportional changes. The Kantian player chooses a representation of the action space, and this choice itself becomes a strategic variable.

At the same time, the analysis reveals an important implementability constraint. When player-specific transformations are permitted, the Kantian condition can be aligned pointwise with the Stackelberg first-order condition along the entire reaction curve. Under strategic complements, this alignment remains compatible with monotone transformations. Under strategic substitutes, however, exact implementation requires one transformation to be decreasing, which violates the Kantian principle of universality. The restriction to a common monotone scale therefore imposes genuine limits on strategic flexibility while preserving the ethical interpretation of Kantian optimization. More broadly, the paper identifies a trade-off between universality and implementability: the stronger the requirement that moral reasoning apply symmetrically across agents, the narrower the set of strategically attainable outcomes.

The economic interpretation of the mechanism is straightforward. In many environments, agents do not merely choose actions; they also choose the metrics through which actions are evaluated. Examples include the normalization of ESG indicators (Christensen et al., 2022), the definition of carbon intensity measures (Hertwich and Peters, 2009), contribution metrics in public-good environments (Andreoni, 1995), or benchmarks for effort and restraint within organizations (Diewert, 1992). Our results show that such choices can affect equilibrium behavior even when they leave underlying material payoffs unchanged. The model therefore provides a formal link between Kantian optimization, framing effects, and institutional design.

While the general results of the paper are cast in terms of stationary and local implementability, the global equilibrium notion introduced in Section 3.5 is not vacuous. For a concretely specified payoff function and an explicitly constructed scale $f$ (rather than merely its local elasticity at two points), the global optimality condition can be verified directly. This is especially relevant when institutional designers or regulators prescribe a complete metric, rather than just its local properties, to evaluate moral behavior.

Several extensions appear promising. Symmetric Kantian populations would generate a meta-game over moral geometry itself, potentially leading to endogenous coordination on strategically advantageous representations. Repeated interactions could make scaling a persistent commitment technology rather than a one-shot strategic choice. The framework may also connect naturally to mechanism design, where a regulator or institution selects the metric through which cooperative behavior is evaluated. More generally, endogenizing the emergence of transformations through bargaining, norms, or political processes could help bridge the gap between strategic representation and the institutional evolution of moral behavior.

## 6. Conclusion

This paper began from a simple but important observation: multiplicative Kantian equilibrium is not strategically invariant under monotone transformations of the strategy space. We asked whether this failure of invariance is merely a technical property of the equilibrium concept or whether it can itself become strategically relevant. Our analysis shows that it can. Once the choice of scale is endogenized, the Kantian player can use the representation of actions to influence equilibrium outcomes.

The central result is a sharp asymmetry between the two canonical strategic regimes. Under strategic substitutes, the Kantian player cannot improve upon the Nash equilibrium payoff, although the appropriate scale can eliminate the losses generated by naive Kantian behavior. In this case, the transformation acts primarily as a defensive instrument. Under strategic complements, by contrast, the Kantian player can implement the Stackelberg leader outcome and obtain a payoff strictly above the Nash benchmark. The strategic value of Kantian behavior therefore depends fundamentally on the structure of interaction in which it is embedded.

More broadly, the paper shows that the parametrization of strategies is not behaviorally neutral in environments based on collective proportional deviations. In Kantian games, equilibrium depends not only on preferences and feasible actions, but also on how coordinated deviations are represented. The geometry of the strategy space affects the equilibrium conditions directly through the scaling structure of collective deviations. As a result, games that are strategically equivalent under Nash behavior need not remain equivalent under Kantian optimization.

The economic implications extend beyond the specific Cournot and Bertrand settings studied here. In many environments, cooperative or morally motivated behavior is expressed through quantitative metrics: emissions standards, ESG indicators, voluntary contribution schemes, productivity targets, or effort norms. In such settings, the definition of the metric may itself influence strategic incentives and equilibrium outcomes. The analysis therefore provides a formal mechanism through which framing and measurement affect cooperation and strategic interaction.

More generally, the paper suggests that the analysis of moral behavior in strategic environments may require attention not only to preferences and incentives, but also to the representation of collective action itself. Understanding how strategic agents coordinate around proportional rules may therefore require studying the endogenous geometry of interaction in addition to standard payoff considerations. We hope the present framework contributes to further research on the institutional and strategic foundations of Kantian cooperation.